\documentclass[titlepage, 12pt, a4paper]{article}
\usepackage[utf8]{inputenc}
\usepackage[english]{babel} 
\usepackage{enumerate}
\usepackage[english]{babel}

\usepackage{amssymb}
\usepackage[numbers]{natbib}
\usepackage[T1]{fontenc}
\usepackage{subfig}
\usepackage{graphicx}
\usepackage{caption}
\usepackage{wrapfig}
\usepackage{float}
\usepackage{url}
\usepackage{csquotes}
\usepackage{caption}
\usepackage{booktabs}
\usepackage{listings}
\usepackage{blindtext}
\usepackage[toc]{appendix}
\usepackage{array, makecell}
\usepackage{amsmath, mathtools}
\usepackage{tikz}

\author{Hyunyeon Kim, Kyung Eun Kim, Soohan Park, and Jongsoo Sohn}
\title{E-voting System Using Homomorphic Encryption and Blockchain Technology to Encrypt Voter Data}

\begin{document}
\maketitle

\tableofcontents
\newpage

\begin{abstract}
Homomorphic encryption and blockchain technology are regarded as two significant technologies for improving e-voting systems. In this paper, we suggest a novel e-voting system using homomorphic encryption and blockchain technology that is focused on encrypting voter data. By encrypting voter information rather than cast votes, the system enables various statistical analyses regarding the vote result while securing the credibility, privacy and verifiability of overall e-voting system. We checked the efficiency of the overall system by comparing the speed of the proposed system with the speed of other e-voting systems that use homomorphic encryption and blockchain technology.

\end{abstract}

\section{Introduction}
The right to vote, one of the most enshrined rights protected by constitutions, has been the fundamental basis of democracy. We must recall that one of the most significant changes the Civil War brought to America was the voting right that male African Americans received \citep{keyssar2009right}\citep{wang1995black}\citep{wang2012trial}. This change is regarded as a major turning point in U.S. History. Similarly, one of the major outcomes of World War 1 was the victory of the women's suffrage movement \citep{ramirez1997changing}; women were finally granted the right to vote after 100 years of struggle \citep{rover2019women}. By winning the right to vote, women and blacks were recognized as \enquote{citizens} of their countries. We can surmise that voting is indispensable for a society to be fully democratic.

Because fair voting is the essence of democracy, the voting system has been established and refined along with the development of democracy. There are some key procedures voting must include: identification and authentication of voters, collecting and recording of votes, vote tallying, and declaration of the results \citep{more2020e2e}. While performing these steps, voting deploys several protocols to ensure eligibility, integrity and auditability. Voter eligibility guarantees that voters who cast votes are eligible. Voting integrity is defined as the global standards voters must follow to cast votes. Auditability refers to the ability of the voting system to conduct reviews after the poll closes. Good auditability ensures accurate counting of votes.

E-voting systems are definitely a better choice than paper ballot systems for ensuring fair voting. It can prevent the wasted time and resources that are inevitable with paper voting systems \citep{lee2006voter}. However, we must also note that e-voting systems display several weaknesses because they rely on internet platforms. The security of the internet platform is crucial to the overall security of an e-voting system. If the former is at risk, the latter is also fatally flawed. In order to reduce this dependence, e-voting systems implement several technologies to protect the creditability of votes and the privacy of voters \citep{jonker2013privacy}. For example, e-voting systems using homomorphic encryption can guarantee the verifiability of ballots and of results \citep{more2020e2e}.

The common goal of paper ballots and e-voting systems is fair voting; votes are counted accurately while the result of each cast vote remains unknown. However, in this paper, we suggest a new e-voting system that encrypts information about the voters rather than the voting ballots. Unlike current e-voting systems that focus on hiding the voting results, this e-voting system focus on voter privacy using homomorphic encryption and blockchain technology. After presenting the background literature in the next section (2), we introduce this novel concept in detail in Section 3. We further analyze the introduced system in Section 4 and present conclusions in Section 5.

\section{Related Works}

\subsection{E-voting System Using Blockchain}
Blockchain is a distributed and decentralized public ledger managed by a peer-to-peer network. Once the data in one block undergoes change, the revision process is recorded and shared by all the other blocks in the blockchain system. In other words, it is impossible to amend the data secretly. This is a clear advantage in an e-voting system because blockchain itself can monitor whether the voting results are manipulated by external forces. Any forced revision of data is detected immediately. The transparency of a blockchain network leads to credibility of the total e-voting system. Moreover, the main characteristic of blockchain system, especially public blockchain, is that its network is decentralized. Decentralized networks avoid reliance on any central authority; decisions for the total system are made by a majority of the members in the network. Decentralization of blockchain networks can prevent any possible corruption of total e-voting systems created by central authorities.

Because the transparency and detectability of a decentralized blockchain network become strengths in an e-voting system, there have been several trials to implement blockchain technology in an e-voting system \citep{dagher2018broncovote}\citep{hanifatunnisa2017blockchain}\citep{hjalmarsson2018blockchain}\citep{khan2018secure} \citep{liu2017voting}\citep{wang2018large}\citep{wu2017voting}. Two in particular, \citep{dagher2018broncovote} and \citep{wu2017voting}, introduced e-voting systems using blockchain and a ring signature. E-voting systems based on blockchain technology might have difficulty in preserving end-to-end privacy. Several cryptographic techniques, such as homomorphic encryption, ring signature, and blind signature, can ensure that a blockchain-based e-voting system is able to preserve the privacy of voters \citep{jonker2013privacy}. In particular, \citep{wang2018large} introduces a new e-voting system that uses blockchain and a ring signature. The uniqueness of this e-voting system is that it enables large-scale elections, unlike other blockchain-based e-voting systems.

\subsection{E-voting System Using Blockchain Technology and Homomorphic Encryption}
\subsubsection{Homomorphic Encryption}
Before going over e-voting systems that use both blockchain and homomorphic encryption, let us look at homomorphic encryption first. Homomorphic encryption is a cryptographic technique that enables computations of encrypted data \citep{gentry2009fully}. To be precise, the result of computing encrypted data is equal to the result of encrypting the computation of plain data \citep{gentry2009fully}. In order to compute encrypted data without using homomorphic encryption, encrypted data must undergo decryption; this process creates a risk of exposing plain data and weakens the overall security of the database. Thus, homomorphic encryption guarantees data security by eliminating the need to send plain data.

Let us note that homomorphic encryption clearly has weaknesses that hinder practical implementation \citep{gentry2011implementing} \citep{naehrig2011can}. Data that deploys homomorphic encryption cannot help but require large capacity; in other words, database systems using homomorphic encryption to reinforce security also requires large data storage. Moreover, the speed of encrypting plain data and of computing encrypted data are not fast enough to be practical. The accuracy of homomorphic encryption for large-scale data is also in doubt because the noise, that occurs with every computation, eventually becomes too large to ignore.

\subsubsection{E-voting System Using Blockchain Technology and Homomorphic Encryption}
Despite these weaknesses, there have been many studies to enable practical implementation of homomorphic encryption \citep{gentry2011implementing}\citep{naehrig2011can}. Homomorphic encryption is especially well-suited with e-voting system because votes can be counted in their \enquote{encrypted} state. Because homomorphic encryption uses a public key to encrypt data into hash numbers, encrypted data can be recorded and stored in a blockchain network. Homomorphic encryption ensures privacy and security while the blockchain database guarantees data integrity and transparency. Thus, homomorphic encryption and a blockchain network complement each other to make a fair e-voting system \citep{balasubramanian2016homomorphic}\citep{more2020e2e}.

\section{E-voting System Focused on Voter Data}

In this section, we propose a novel e-voting system that implements blockchain technology and a homomorphic encryption scheme as the basis of a new paradigm in voting systems. Normally, e-voting systems focus on preserving the privacy of cast votes to ensure the credibility of the system. However, the focus of the proposed system is encrypting voter data. By implementing homomorphic encryption, we can perform statistical analysis on voter information while protecting voter privacy. Note that this new approach can complement the weakness of previous e-voting systems using homomorphic encryption. When cast votes undergo homomorphic encryption, voters cannot guarantee whether the cipher texts are equal to the votes they cast. However, because the new e-voting system encrypts only the voter data rather than the votes cast, voters can check their cast votes in the form of plain text. Moreover, because homomorphic encryption allows SQL query on the encrypted data \citep{gahi2015secure}, the analysis can be much more complicated and detailed. We believe that this e-voting system could be a new method by which to interpret elections results from a variety of angles.

\subsection{Proposed system}

In order to fetch information from voters, we first need to choose which information we want to collect. The figure below shows several criteria by which to perform statistical analysis on voters.

\begin{figure}[H]
\begin{center}
\includegraphics[scale=0.4]{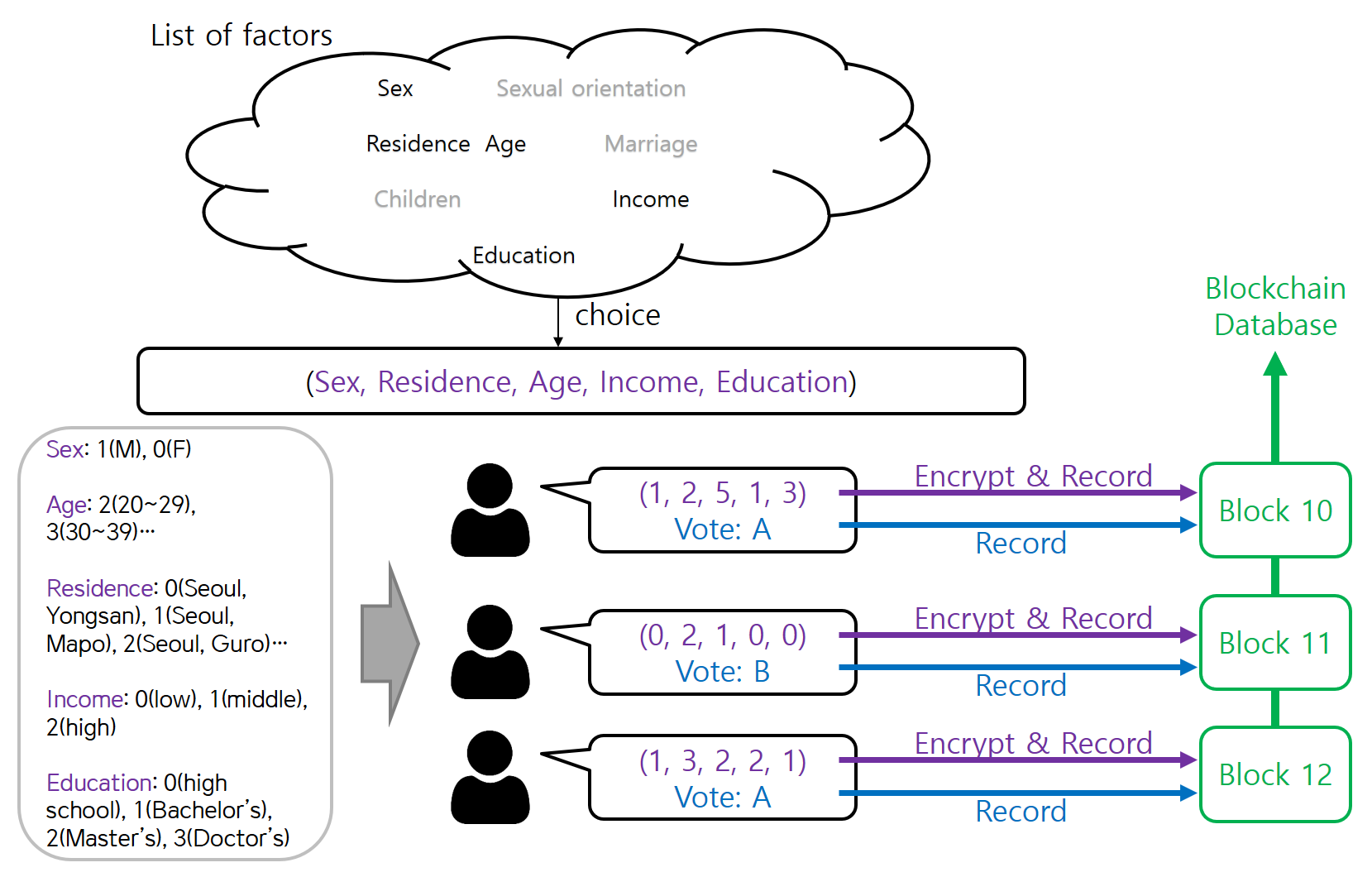} 
\caption{Choosing several features to perform statistical analysis.} \label{1}
\end{center}
\end{figure}

In Figure \ref{1}, there are several attributes that could be statistical criteria. Based on the aim and objective of the vote, we can choose specific factors that might be useful for interpreting the voting result. For example, we choose five traits: sex, residence, age, income and education in Figure \ref{1}. These five features are irreplaceable characteristics for explaining South Koreans. Each voter would give replies to questions regarding these factors. The grey box in Figure \ref{1} explains how the factors can be expressed in terms of an integer. Let us note that each five integers are merged into a batch that does not include \enquote{the voting result of the voter}. The batch from each voter undergoes homomorphic encryption while the voting result remains plain text. 
Encrypted batches and voting results are then recorded on assigned blocks in the blockchain network.

\begin{figure}[H]
\begin{center}
\includegraphics[scale=0.8]{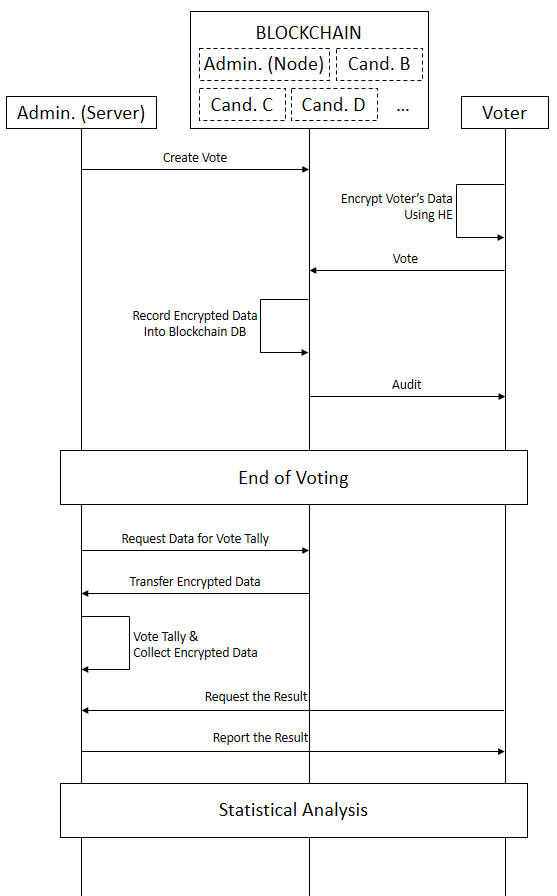} 
\caption{Overall concept of the e-voting system} \label{2}
\end{center}
\end{figure}

Figure \ref{2} depicts the overall progress of an e-voting system focused on voter information. Let us check the three sides that comprise the new e-voting system using blockchain technology and homomorphic encryption: administration server, blockchain system and voters.
We can consider the \enquote{administration server} as the election commission. This server creates votes, counts votes, collects encrypted data, and reports the results to voters via transferring data to the blockchain system.
The \enquote{blockchain network} in this e-voting system, contains several nodes: administration node and nodes assigned to each candidate. This network records encrypted and plain data and transfer data from voters to server and from sever to voters.
A \enquote{voter} in this e-voting system uses a mobile device for e-voting. These devices encrypt voter data using homomorphic encryption and send the data to the blockchain network. We must note that the votes cast are also passed on to the blockchain network in the form of plain text. After the vote tally, participating devices send requests to an administration server to receive the voting result.

Now, let us highlight the important work flows of this new e-voting system. After collecting voter information following the process depicted in Figure \ref{1}, the voter data and cast votes are recorded on assigned blocks in the blockchain system. As the voter information undergoes homomorphic encryption, it can be counted and computed along with the vote tally. Last, we perform statistical analyses after reporting the voting results to voters.

\subsection{Voting Procedure in the Proposed system}

The proposed system consists of the following six modules which express the overall voting procedure. In this section, we introduce each module to explain its function.\\

\textbf{1) Module 1: verify\_voter}

In this first module, we check whether the voter has the right to cast a vote in the e-voting system. In other words, this module performs identification and authentication of voters. Each voter sets an ID and password to login to the e-voting system whenever they want. Let us note that the voters can later check their cast vote by re-accessing the system. \\

\textbf{2) Module 2: collect\_voter}

After choosing factors that work as criteria for later analysis, the administration server asks questions to voters regarding these attributes. This module gathers information from voters, encrypts the collected data using homomorphic encryption, then records the encrypted replies on the blockchain database. \\

\textbf{3) Module 3: cast\_vote}

In this module, voters cast their votes. These cast votes do not undergo homomorphic encryption; votes are recorded in the blockchain network in the form of plain text. \\

\textbf{4) Module 4: check\_vote}

This module checks the vote result of each voter just in case the voter accidentally picks a different choice or the vote itself is wrongly recognized. Each voter has set an ID and password in module \{verify\_vote\}. Voters can use their ID and password to login to the system to check their votes. \\

\textbf{5) Module 5: tally\_vote}

Because cast votes are in plain text form, every vote can be recorded and counted simultaneously. When all the votes are counted, the administration server announces the result to the voters. \\

 \textbf{6) Module 6: analyze\_voters}

In this last module, we perform statistical analysis of the voter information to gain various interpretations of the voting result. The factors chosen above may work as criteria of the statistical analysis. We must note that all these information about the voter remains encrypted. The data is decrypted only when the analysis result comes out.

\section{Results and Discussion}

Let us now compare this newly proposed e-voting system with other e-voting systems that use homomorphic encryption and blockchain technology.

The main and definite obstacle to homomorphic encryption to be used commercially would be \enquote{speed}. Thus, we compared the systems based on how fast the vote result comes out.

In Table \ref{3}, we can check specific settings of the e-voting systems. We use SEAL as a homomorphic encryption library and Hyperledger Fabric 1.4 for the blockchain system as common settings. We also set the total number of voters to 1024. The specific reason for using SEAL is that SEAL is faster than any other homomorphic encryption library because its computations are decimal not binary \citep{melchor2018comparison}.

\begin{table}[H]
\begin{center}
\begin{tabular}{ |m{3cm} || m{6cm} | m{5cm}| }
\hline
 & \makecell{Newly proposed e-voting system} & \makecell{Other systems using HE \\ and blockchain technology}
\\ \hline \hline
\makecell{Encrypted Data} & \makecell{Voter Information} & \makecell{Votes}
\\ \hline
 \makecell{Data Size} & \makecell{400 Bytes(5 Factors) \\ + 1 Byte(data size of each vote)} & \makecell{80 Bytes}
\\ \hline
\makecell{HE addition \\ in vote counting} & \makecell{X} & \makecell{O}
\\ \hline \hline
\multicolumn{3}{|l|}{

}
\\
\multicolumn{3}{|l|}{
$\bullet$ Homomorphic Encryption Library: SEAL
}
\\
\multicolumn{3}{|l|}{
$\bullet$ Blockchain System: Hyperledger Fabric 1.4 \citep{hyperledger}
}
\\
\multicolumn{3}{|l|}{
$\bullet$ Number of voters: 1024
}
\\
\multicolumn{3}{|l|}{

}
\\ \hline
\end{tabular}
\caption{Comparison of the new e-voting system with another e-voting system that uses the HE and the blockchain technology} \label{3}

\end{center}
\end{table}

As shown in Table \ref{3}, the newly proposed e-voting system encrypts voter information while other systems encrypt the votes cast. In other words, the new system records voter information as cipher text and cast votes as plain text while others records cast votes as cipher text. This difference leads to different data size of each block in blockchain systems. For other e-voting systems, each block records only one cipher text, the encrypted data of the cast vote. The data size of this single cipher text is 80 bytes.

\begin{figure}[H]
\begin{center}
\includegraphics[scale=0.8]{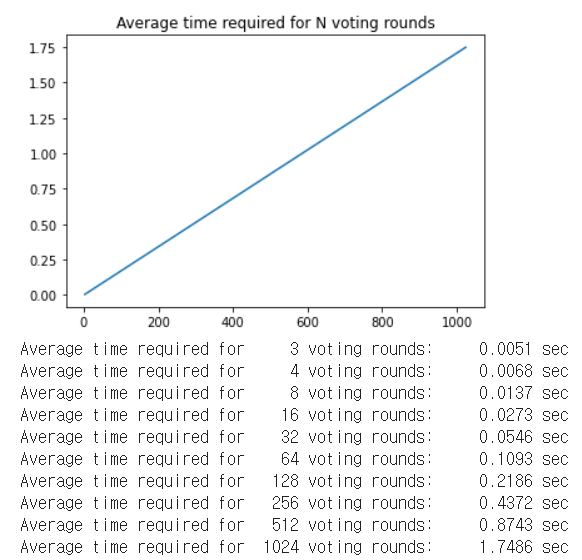}
\includegraphics[scale=0.8]{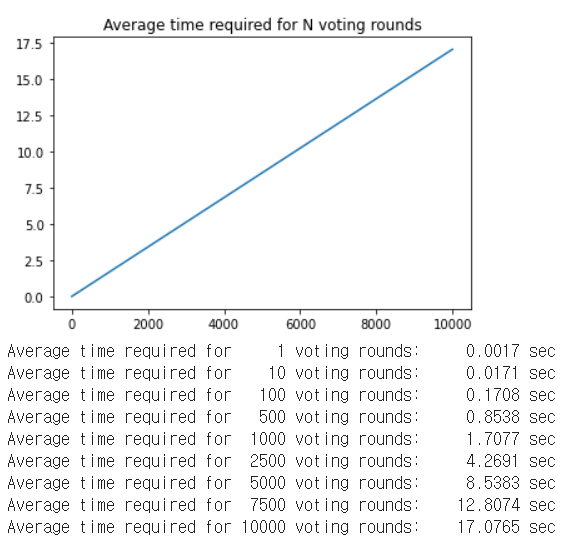}
\caption{Average time required for N voting rounds with data size 400+ bytes (top) and with data size 80 bytes (bottom).\label{4}}
\end{center}
\end{figure}

As for the newly proposed e-voting system, each block contains five cipher texts because there are five factors, and the total data size of five cipher texts is 400 bytes. In addition, we must also count the data size of the cast votes. The data size of each vote may vary depending on the data type. If the data type is integer, the data size would be 4 bytes; if the data type is a single capital alphabet, the data size would be 1 byte. In Table \ref{3}, the data size of each vote is set to 1 byte. However, we must note that the data size of each vote does not matter (see the graph in Figure \ref{4}).

Both of the y-axes of the two graphs in Figure \ref{4} represent the number of voters while the x-axes show the estimated time. 
The top graphic in Figure \ref{4} shows the average time required for N voting rounds with data size over 400 bytes while the bottom graphic shows the average time required for N voting rounds with data size of 80 bytes. For 1024 voters, both graphs show the estimated time of 1.7 seconds.
We can check that the average time estimated for N voting rounds remains the same regardless of data size.

Another major difference would be the time required for counting votes. This difference is based on the existence and non-existence of additional evaluation of the encrypted data. As shown in Table \ref{3}, the newly proposed e-voting system does not require additions of encrypted data while other systems do. For the new e-voting system, computations of encrypted data are required only for statistical analyses performed after the vote counting process. Statistical analyses may take time as they require computations of encrypted data; however, this does not matter because the analyses are carried out after the vote counting.

\begin{figure}[H]
\begin{center}
\includegraphics[scale=1.2]{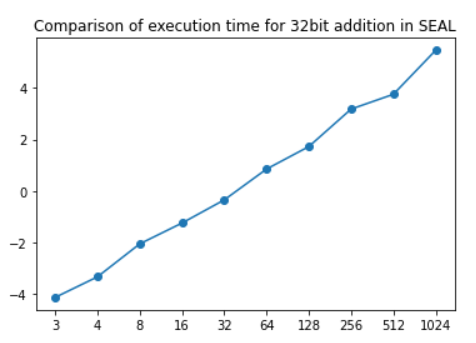} 
\caption{Execution time for 32-bit addition in SEAL} \label{5}
\end{center}
\end{figure}

In Figure \ref{5}, we can check that the estimated time for 1024 additions of encrypted data is 5.463. On the other hand, counting votes in plain text does not take any more time because the votes are recorded and counted simultaneously. Thus, the time difference of the vote counting process between the new e-voting system and others is never significant. We proved that the e-voting system that focuses on encrypting voter data is at least as efficient as the other e-voting systems using homomorphic encryption and blockchain technology.

\section{Conclusions}

E-voting systems have undergone significant developments by implementing homomorphic encryption and blockchain technology. Homomorphic encryption guarantees ballot verifiability and result verifiability of votes while the blockchain technology promises integrity and transparency. However, these e-voting systems using homomorphic encryption and blockchain technology do show weakness. Because the cast votes are encrypted in the form of a cipher text, voters cannot be certain that the cast votes are the same as the encrypted forms.

In this paper, we suggest a novel e-voting system using homomorphic encryption and blockchain technology that focuses on encrypting voter information rather than the cast votes. Cast votes remain in plain text while only voter data is encrypted; thus, voters can always check their cast votes. The main advantage of this new e-voting system is that it enables statistical analyses of the vote results while protecting voter privacy. In order to perform statistical analysis in detail, the e-voting system may ask questions to voters regarding several attributes, such as age, gender, and education. Replies to these questions surely contain private information about voters; however, voters can answer with no worries of exposing themselves because every reply undergoes homomorphic encryption. The collected information need not be decrypted even during the statistical analyses, highlighting the strength of homomorphic encryption.

Because speed is the main factor for successful commercialization of homomorphic encryption, we checked the efficiency of this new e-voting system by comparing the speed of vote counting compared to the speed in other systems. The speed of recording data in the blockchain system does not differ by data size. Although the new e-voting system requires larger data size for each block compared to other systems, there is no difference in the recording speed. However, there exists a difference in speed during the vote counting process. The new system counts votes in the form of plain text while other systems counts votes that are in encrypted forms. In other words, other systems require additions of encrypted data while the new system does not. Thus, the speed of vote counting in the new system is faster than the speed of vote counting in the other e-voting systems using blockchain technology and homomorphic encryption. Let us note that the new e-voting system, which encrypts voter data, might require multiplication and addition of encrypted data in statistical analyses. Because the statistical analyses are performed after the votes are counted, the actual speed of getting the voting result does not depend on the computation speeds for homomorphic encryption.

We believe that this e-voting system offers a new paradigm for voting systems. This system could enable further statistical analyses while complementing the weakness of previous e-voting systems using homomorphic encryption and blockchain technology. Moreover, this system displays efficiency at least as high as other systems do. Thus, we hope that this new system brings meaningful change to e-voting and can be utilized in numerous fields.

\newpage

%next line adds the Bibliography to the contents page
\addcontentsline{toc}{section}{Bibliography}
%uncomment next line to change bibliography name to references
%\renewcommand{\bibname}{References}
\bibliographystyle{plain}
\bibliography{ref}

\begin{thebibliography}{10}

\bibitem{balasubramanian2016homomorphic}
Kannan Balasubramanian, M~Jayanthi, et~al.
\newblock A homomorphic crypto system for electronic election schemes.
\newblock {\em Circuits and Systems}, 7(10):3193, 2016.

\bibitem{dagher2018broncovote}
Gaby~G Dagher, Praneeth~Babu Marella, Matea Milojkovic, and Jordan Mohler.
\newblock Broncovote: Secure voting system using ethereum’s blockchain.
\newblock 2018.

\bibitem{gahi2015secure}
Youssef Gahi, Mouhcine Guennoun, and Khalil El-Khatib.
\newblock A secure database system using homomorphic encryption schemes.
\newblock {\em arXiv preprint arXiv:1512.03498}, 2015.

\bibitem{gentry2009fully}
Craig Gentry.
\newblock {\em A fully homomorphic encryption scheme}.
\newblock Stanford university, 2009.

\bibitem{gentry2011implementing}
Craig Gentry and Shai Halevi.
\newblock Implementing gentry’s fully-homomorphic encryption scheme.
\newblock In {\em Annual international conference on the theory and
  applications of cryptographic techniques}, pages 129--148. Springer, 2011.

\bibitem{hanifatunnisa2017blockchain}
Rifa Hanifatunnisa and Budi Rahardjo.
\newblock Blockchain based e-voting recording system design.
\newblock In {\em 2017 11th International Conference on Telecommunication
  Systems Services and Applications (TSSA)}, pages 1--6. IEEE, 2017.

\bibitem{hjalmarsson2018blockchain}
Fri{\dh}rik~{\TH} Hj{\'a}lmarsson, Gunnlaugur~K Hrei{\dh}arsson, Mohammad
  Hamdaqa, and G{\'\i}sli Hj{\'a}lmt{\`y}sson.
\newblock Blockchain-based e-voting system.
\newblock In {\em 2018 IEEE 11th International Conference on Cloud Computing
  (CLOUD)}, pages 983--986. IEEE, 2018.

\bibitem{jonker2013privacy}
Hugo Jonker, Sjouke Mauw, and Jun Pang.
\newblock Privacy and verifiability in voting systems: Methods, developments
  and trends.
\newblock {\em Computer Science Review}, 10:1--30, 2013.

\bibitem{keyssar2009right}
Alexander Keyssar.
\newblock {\em The right to vote: The contested history of democracy in the
  United States}.
\newblock Basic Books, 2009.

\bibitem{khan2018secure}
Kashif~Mehboob Khan, Junaid Arshad, and Muhammad~Mubashir Khan.
\newblock Secure digital voting system based on blockchain technology.
\newblock {\em International Journal of Electronic Government Research
  (IJEGR)}, 14(1):53--62, 2018.

\bibitem{lee2006voter}
Kwang-Woo Lee, Yun-Ho Lee, Dong-Ho Won, and Seung-Joo Kim.
\newblock A voter verifiable receipt in electronic voting with improved
  reliability.
\newblock {\em Journal of the Korea Institute of Information Security \&
  Cryptology}, 16(4):119--126, 2006.

\bibitem{liu2017voting}
Yi~Liu and Qi~Wang.
\newblock An e-voting protocol based on blockchain.
\newblock {\em IACR Cryptol. ePrint Arch.}, 2017:1043, 2017.

\bibitem{melchor2018comparison}
Carlos~Aguilar Melchor, Marc-Olivier Kilijian, C{\'e}dric Lefebvre, and Thomas
  Ricosset.
\newblock A comparison of the homomorphic encryption libraries helib, seal and
  fv-nfllib.
\newblock In {\em International Conference on Security for Information
  Technology and Communications}, pages 425--442. Springer, 2018.

\bibitem{more2020e2e}
Pranav More, Rohit Nawale, Reena Kharat, Mohit Nakhale, and Kaustubh Patil.
\newblock E2e verifiable blockchain voting system using homomorphic encryption.
\newblock {\em Journal of Critical Reviews}, 7(19):1120--1127, 2020.

\bibitem{naehrig2011can}
Michael Naehrig, Kristin Lauter, and Vinod Vaikuntanathan.
\newblock Can homomorphic encryption be practical?
\newblock In {\em Proceedings of the 3rd ACM workshop on Cloud computing
  security workshop}, pages 113--124, 2011.

\bibitem{hyperledger}
Hyperledger Performance and Scale~Working Group.
\newblock Hyperledger blockchain performance metrics.
\newblock 2018.

\bibitem{ramirez1997changing}
Francisco~O Ramirez, Yasemin Soysal, and Suzanne Shanahan.
\newblock The changing logic of political citizenship: Cross-national
  acquisition of women's suffrage rights, 1890 to 1990.
\newblock {\em American sociological review}, pages 735--745, 1997.

\bibitem{rover2019women}
Constance Rover.
\newblock {\em Women's Suffrage and Party Politics in Britain, 1866--1914}.
\newblock University of Toronto Press, 2019.

\bibitem{wang2018large}
Baocheng Wang, Jiawei Sun, Yunhua He, Dandan Pang, and Ningxiao Lu.
\newblock Large-scale election based on blockchain.
\newblock {\em Procedia Computer Science}, 129:234--237, 2018.

\bibitem{wang1995black}
Xi~Wang.
\newblock Black suffrage and the redefinition of american freedom, 1860-1870.
\newblock {\em Cardozo L. Rev.}, 17:2153, 1995.

\bibitem{wang2012trial}
Xi~Wang.
\newblock {\em The trial of democracy: Black suffrage and northern Republicans,
  1860-1910}.
\newblock University of Georgia Press, 2012.

\bibitem{wu2017voting}
Yifan Wu.
\newblock An e-voting system based on blockchain and ring signature.
\newblock {\em Master. University of Birmingham}, 2017.

\end{thebibliography}

\end{document}